\documentstyle[11pt,fleqn]{article}

\def\beq{\begin{equation}}
\def\eeq{\end{equation}}
\def\bea{\begin{array}}
\def\ea{\end{array}}
\def\beqn{\begin{eqnarray}}
\def\eeqn{\end{eqnarray}}
\def\ni{\noindent}
\def\a{\alpha}

\def\L{\Lambda}

\def\o{\omega}
\def\e{\epsilon}

\def\S{\Sigma}

\def\Uqso{U'_q({\rm so}_n)}

\begin{document}

\ni
{\LARGE\bf The Use of Quantum Algebras in}\\
\vspace{-2.4mm}

\ni
{\LARGE\bf Quantum Gravity}

\vspace{5mm}
\ni
{\sl A. M.~GAVRILIK}\footnote{omgavr@bitp.kiev.ua}

\vspace{2.8mm}
\ni
{\it Bogolyubov Institute for Theoretical Physics, \\
Metrologichna Street 14b,  03143 Kyiv, \ Ukraine  }

\vspace{1mm}
\begin{abstract}
\small
After a brief survey of the appearance
of quantum algebras in diverse contexts of quantum gravity,
we demonstrate that the particular deformed algebras,
which arise within the approach of J.Nelson and T.Regge
to 2+1 anti-de Sitter quantum gravity (for space surface
of genus $g$) and which are basic for generating the algebras
of independent quantum observables, are in fact isomorphic to the
nonstandard $q$-deformed analogues $U'_q(so_n)$
(introduced in 1991) of Lie algebras of the orthogonal
groups $SO(n)$, with $n$ linked to $g$ as $n=2g+2$.
\end{abstract}

\vspace{3mm}

\ni
{\large\bf 1. Introduction}

\vspace{1.2mm}
Quantum or $q$-deformed algebras may appear in quantum
(or $q$-versions of) gravity in various situations. Let us mension
some of them.

\vspace{1.2mm}
$\bullet$ Case of $n$ spacetime dimensions ($n\ge 2$), straightforward
  approach to construct $q$-gravity (this is accomplished, e.g.,
in \cite{Cas}). Basic steps are:

- Start with some version of quantum/$q$-deformed algebra $iso_q(n)$
  (in \cite{Cas} it is projected out from the standard quantum algebras
  $U_q(B_r)$ or $U_q(D_r)$ of Drinfeld and Jimbo \cite{DJ}). In the
  particular Poincare algebra $iso_q(3,1)$ exploited by Castellani,
  only those commutation relations which involve momenta do depend
  on the parameter $q$,
  while the Lorentz subalgebra remains non-deformed;

- Develop necessary bicovariant differential calculus;

- A $q$-gravity is constructed by "gauging" the $q$-analogue of
  Poincare algebra. The resulting Lagrangian turns out to be a
  generalization \cite{Cas} (see also \cite{Bi}) of the usual
  Einstein's or Einstein-Cartan's one.

It is worth to empasize that in this approach the obtained results,
including physical implications, unambiguously depend on the specific
features of chosen $q$-algebra.

\vspace{1.2mm}
$\bullet$ Two-dimensional quantum Liouville gravity \cite{JaP},
 within particular framework of quantization, leads to the appearance
 \cite{GN} of quantum algebras such as $U_q(sl(2,{\bf C}))$.

\vspace{1.2mm}
$\bullet$ Case of $3$-dimensional (Euclidean) gravity. The simplicial
  approach developed by Ponzano and Regge \cite{PR} employs
  irreducible representations of the algebra $su(2)$
  labelled by spins $j$ and assigned to edges of tetrahedra
  in triangulation, the main ingredient being
  $6j$-symbols of $su(2)$. Within natural generalization
  of this approach by Turaev and Viro \cite{TV}, see also \cite{MT},
  the underlying symmetry of the action
  (which can be related to Chern-Simons theory)
  is that of the quantum algebra $su_q(2)$,
  and basic objects are $q-6j$ symbols. Due to this,
  physical quantities become expressible through
  topological (knot or link) invariants.
  The parameter $q$ takes into account cosmological constant
  and, on the other hand, is connected with
  the (quantized) Chern-Simons coupling
  constant $k\ $  as $\ q=\exp \frac{2i \pi}{k+2}$.

\vspace{1.2mm}
$\bullet$ ($2+1$)-dimensional gravity with or without
  cosmological constant $\L$ is known to possess important
  peculiar features \cite{DJt,Wi}.
  Within the approach to quantization developed by J.~Nelson
  and T.~Regge,  specific deformed algebras arise \cite{NRZ,NR}
  for the situation with $\L < 0$, and just
  this fact will be of our main concern here.

\vspace{3mm}
\ni {\large\bf 2. Nonstandard $q$-deformed algebras $\Uqso$,
                 their advantages}
\vspace{2mm}

As defined in \cite{GK}, the nonstandard $q$-deformation $\Uqso$
of the Lie  algebra ${\rm so}_n$ is
given as a complex associative algebra with ${n-1}$
generating elements $I_{21}$, $I_{32}, \ldots$, $I_{n,n-1}$
obeying the defining relations (denote $q+q^{-1}\equiv [2]_q$)
\beq
\begin{array}{l}
I_{j,j-1}^2I_{j-1,j-2} + I_{j-1,j-2}I_{j,j-1}^2 -
[2]_q \ I_{j,j-1}I_{j-1,j-2}I_{j,j-1} = -I_{j-1,j-2}, \\[2mm]
I_{j-1,j-2}^2I_{j,j-1} + I_{j,j-1}I_{j-1,j-2}^2 -
[2]_q \ I_{j-1,j-2}I_{j,j-1}I_{j-1,j-2} = -I_{j,j-1},  \\[2mm]
[I_{i,i-1},I_{j,j-1}] =0  \qquad {\rm if} \quad \mid {i-j}\mid >1.
\end{array}  \label{f1}
\eeq
At $q\to 1$, $\ [2]_q\to 2$ (non-deformed or classical limit),
these go over into defining relations of the $so(n)$ Lie algebras.

Among the {\it advantages of these nonstandard
$q$-deformed algebras} with regards to the Drinfeld-Jimbo
quantum deformations, the following should be pointed out.

({\bf i}) Existence
of the canonical chain of embedded subalgebras
(from now on, we omit the prime in the symbol)
\[
U_q(so_{n})\supset U_q(so_{n-1})\supset \cdots \supset
U_q(so_{4})\supset U_q(so_{3})
\]
in the case of $U_q(so_n)$ and, due to this, implementability
of the $q$-analogue of Gelfand-Tsetlin formalism enabling one
to construct finite dimensional representations \cite{GK,GI}.

({\bf ii}) Existence, for all the real forms known in the
nondeformed case $q=1$, of their respective $q$-analogues --
the  "compact" $U_q(so_n)$ and the "noncompact" $U_q(so_{p,s})$
 (with $p+s=n$) real forms. Moreover, each such form exists along
 with the corresponding chain of embeddings.
 For instance, in the $n$-dimensional $q$-Lorentz case we have
\[
U_q(so_{n-1,1})\supset U_q(so_{n-1})\supset U_q(so_{n-2})\supset \cdots
\supset U_q(so_{3}).
\]
This fact allows us to develop the construction and analysis of
infinite dimensional representations of
$U_q(so_{n-1,1})$, see \cite{GK,GK2}.

({\bf iii}) Existence of embedding
$U_q(so_{3})\subset U_q(sl_{3})$ generalizable \cite{Nou} to the
embedding of higher $q$-algebras such that
$U_q(so_{n})\subset U_q(sl_{n})$, -- the fact which
enables construction of the proper quantum analogue \cite{Nou}
of symmetric coset space $SL(n)/SO(n)$.

({\bf iv}) If one attempts to get a $q$-analogue of the Capelli
identity known to hold for the dual pair $sl_2 \leftrightarrow so_n$,
nothing but the nonstandard $q$-algebra $U_q(so_n)$ given in (1)
inevitably arises \cite{NUW}.
As a result, the relation $Casimir\{U_q(sl_2)\}=Casimir\{U_q(so_n)\}$
is valid \cite{NUW,NUW2} within particular representation.

({\bf v}) Natural appearance, as will be discussed in Sec.4, of these
$q$-algebras within the Nelson-Regge approach to $2+1$ quantum gravity.

As a drawback, let us mention the fact that Hopf algebra structure
is not known for $U_q(so_n)$, although for the situation ({\bf iii})
the nonstandard $q$-algebra $U_q(so_n)$ was shown to be a coideal
\cite{Nou} in the Hopf algebra $U_q(sl_n)$.

Recall that it was ({\bf i}), ({\bf ii}) which motivated
introducing in \cite{GK} this class of $q$-algebras.

\vspace{3mm}
\ni {\large\bf 3. Bilinear formulation of $U_q(so_n)$ }
\vspace{2mm}

Along with the definition in terms of trilinear relations (1) above,
a {\it `bilinear' formulation} of $U_q(so_n)$ can as well be
provided.  To this end, one introduces the generators
(set $k > l+1, \ \  1\leq k,l \leq n$)
\[
I^{\pm}_{k,l}\equiv [I_{l+1,l} , I^{\pm}_{k,l+1}]_{q^{\pm 1}}
\equiv q^{\pm 1/2}I_{l+1,l} I^\pm_{k,l+1} -
q^{\mp 1/2}I^\pm_{k,l+1} I_{l+1,l}
\]
together with $I_{k+1,k} \equiv I^+_{k+1,k}\equiv I^-_{k+1,k}$.
Then (\ref{f1}) imply
\[
[I^+_{lm} , I^+_{kl}]_q = I^+_{km}, \ \ \
[I^+_{kl} , I^+_{km}]_q = I^+_{lm}, \ \ \
[I^+_{km} , I^+_{lm}]_q = I^+_{kl}  \ \ \ {\rm if} \ \ k>l>m,
\]
\beq                                                      \label{f2}
[I^+_{kl} , I^+_{mp}] = 0 \ \ \ \ \ \ {\rm if} \ \ \ k>l>m>p
\ \ \ \ \ \ {\rm or \ \ \ \ if} \ \
k>m>p>l;
\eeq
\[
[I^+_{kl} , I^+_{mp}] = (q-q^{-1}) (I^+_{lp}I^+_{km}-
I^+_{kp}I^+_{ml}) \ \ \ \ \ \ {\rm if} \ \ \ k>m>l>p.
\]
Analogous set of relations exists which involves $I_{kl}^-$ along with
$q\to q^{-1}$ (denote this ``dual'' set by (\ref{f2}$'$)).
In the `classical' limit $q\to 1$ , both (\ref{f2}) and
(\ref{f2}$'$) reduce to those of ${\rm so}_n$.

To illustrate, we give the examples of $n=3$,
isomorphic to Fairlie -- Odesskii algebra \cite{OF},
and $n=4$ (recall that the $q$-commutator is defined as
$[X,Y]_q\equiv q^{1/2}X Y - q^{-1/2} Y X$):
\[ \hspace{-0.2cm}
U_q({\rm so}_4) \hspace{0.2cm}
\left\{ \hspace{0.2cm}
\vspace{-2mm}
\parbox{12cm}
{
\beq
\hspace{-.8cm}
U_q({\rm so}_3): \hspace{0.6cm}
[I_{21} , I_{32}]_q = I_{31}^+, \ \
[I_{32} , I_{31}^+]_q = I_{21}, \ \
[I_{31}^+ , I_{21}]_q = I_{32}.                       \label{FO}
\eeq
\beq
\hspace{-1.0cm}
\begin{array}{cclll}

 \hspace{0.4cm}& \hspace{0cm}&
[I_{32},I_{43}]_q = I_{42}^+,\ \ &
[I_{31}^+,I_{43}]_q = I_{41}^+,\ \ &
[I_{21},I_{42}^+]_q = I_{41}^+, \\
&&
[I_{43},I_{42}^+]_q=I_{32},&
[I_{43},I_{41}^+]_q=I_{31}^+,&
[I_{42}^+,I_{41}^+]_q=I_{21}, \\
&&
[I_{42}^+,I_{32}]_q=I_{43},&
[I_{41}^+,I_{31}^+]_q=I_{43},&
[I_{41}^+,I_{21}]_q=I_{42}^+,
\end{array}                                             \label{O4}
\eeq
\vspace{-0.4cm}

\beq
\hspace{-1.0cm}
[I_{43},I_{21}]=0,\ \  [I_{32},I_{41}^+]=0,\ \
[I_{42}^+,I_{31}^+]=(q-q^{-1})(I_{21}I_{43}-I_{32}I_{41}^+).   \label{O4p}
\eeq
}
\right.
\]
The first relation in (\ref{FO}) is viewed as definition
for third generator $I_{31}^+$; with this,
the algebra is given in terms of $q$-commutators. Dual copy
of $U_q({\rm so}_3)$ involves the generator
$I_{31}^-=[I_{21},I_{32}]_{q^{-1}}$ which enters the
relations same as (\ref{FO}), but with $q\to q^{-1}$.
Similar remarks concern the generators $I_{42}^+$, $I_{41}^+$,
as well as (dual copy of) the whole algebra $U_q({\rm so}_4)$.

\vspace{3mm}
\ni {\large\bf 3. The deformed algebras $A(n)$ of Nelson and Regge}
\vspace{2mm}

For $(2+1)$-dimensional gravity with cosmological constant $\L < 0$,
the lagrangian density involves spin connection $\o_{ab}$ and
dreibein $e^a$, $\ a,b=0,1,2$, combined in the $SO(2,2)$-valued
(anti-de Sitter) spin connection $\o_{AB}$ of the form
\[   \hspace{28mm}
\o_{AB} = \left(
\bea{ll}
       \o_{ab}  &  \frac1\a e^a \\
       -\frac1\a e^b &  0
\ea
\right) ,
\]
and is given in the Chern-Simons (CS) form \cite{Wi}
\[
\frac{\a}{8} (\hbox{d} \o^{AB} -
   \frac23 \o^{A}_{\ F}\wedge \o^{FB} )\wedge \o^{CD} \e_{ABCD}.
\]
Here $A,B=0,1,2,3\ $, the metric is $\eta_{AB} = (-1,1,1,-1) $,
and the CS coupling constant is connected with $\L$, so that
$\L = - \frac{1}{3\a^2}$.
The action is invariant under $SO(2,2)$, leads to Poisson brackets
and field equations. Their solutions, i.e. infinitesimal connections,
describe space-time which is locally anti-de Sitter.

To describe global features of space-time,
within fixed-time formulation, of principal importance
are the {\it integrated connections} which provide a mapping $S:$
$\pi_1(\S )\to G$ of the homotopy group for a space surface $\S$
into the group $G= SL_+(2,R)\otimes SL_-(2,R)$
(spinorial covering of $SO(2,2)$) and thoroughly
studied in \cite{NRZ}.
To generate the algebra of observables, one takes the traces
\[
c^{\pm}(a)= c^{\pm}(a^{-1}) =\frac12 {\rm tr} [S^{\pm}(a)],\ \ \ \
a\in \pi_1, \ \ \ \ S^{\pm }\in SL_{\pm}(2,R).
\]
For $g=1$ (torus) surface $\S$, the algebra of (independent)
quantum observables has been derived \cite{NRZ}, which turned out
to be isomorphic to the cyclically symmetric Fairlie -- Odesskii
algebra \cite{OF}. This latter algebra, however, is known to coincide
\cite{GK2} with the special $n=3$ case of $U_q(so_n)$. So, natural
question arises whether for surfaces of higher genera $g\ge 2$,
the nonstandard $q$-algebras $U_q(so_n)$ also play a role.

Below, the positive answer to this question is given.

For the topology of spacetime $\S\times{\bf R}$
($\S$ being genus-$g$ surface), the homotopy group $\pi_1(\S)$
is most efficiently described in terms of $2g+2=n$ generators
$t_1, t_2, \ldots, t_{2g+2}$ introduced in \cite{NR} and
such that
\[
t_1 t_3 \cdots t_{2g+1} = 1,      \hspace{12mm}
t_2 t_4,..., t_{2g+2} = 1,  \hspace{10mm} \hbox{and}
\hspace{8mm}    \prod_{i=1}^{2g+2} t_i =1.
\]
\vspace{2mm}

\ni
Classical gauge invariant trace elements ($n(n-1)/2$ in total)
defined as
\beq
\a_{ij} = \frac12 {\rm Tr} (S(t_i t_{i+1} \cdots t_{j-1})), \ \ \ \
                 S\in SL(2,R),
                                               \label{clas}
\eeq
generate concrete algebra with Poisson brackets, explicitly found
in \cite{NR}. At the quantum level, to the algebra with generators
(\ref{clas}) there corresponds quantum commutator algebra $A(n)$
specific for $2+1$ quantum gravity with negative 
$\L$. For each quadruple of indices $\{j,l,k,m\},\ j,l,k,m=1,\ldots,n,$
obeying (see \cite{NR}) `anticlockwise ordering'
\vspace{0.1mm}
\beq
\hspace{52mm}
{  \scriptsize
\begin{array}{lcl}
               {}   &  j   &  {}        \\
         \swarrow   &  {}  &  \nwarrow  \\
        \hspace{-2.6mm}l \ \ \ \ &  {}  &  \ \ \  m  \ \ ,\\
          \searrow  &  {}  &  \nearrow  \\
               {}   &  k   &  {}        \\
\end{array}                      }    \label{four}
\eeq
\vspace{0.1mm}
\renewcommand{\normalsize}
\ni the algebra $A(n)$ of quantum observables reads \cite{NR}:
\beq                                                      \label{An}
\begin{array}{clc}
    {} &  [a_{mk}, a_{jl}] = [a_{mj}, a_{kl}] = 0 , &  {}
          \vspace{2mm}     \\

{}  & [a_{jk}, a_{kl}] = (1-\frac1K) (a_{jl} - a_{kl} a_{jk}) , &  {}
\vspace{2mm}      \\
  {}  & [a_{jk}, a_{km}] = (\frac1K -1) (a_{jm} - a_{jk} a_{km}) , &  {}
\vspace{2mm}      \\
  {}  & [a_{jk}, a_{lm}] = (K-\frac1K) (a_{jl} a_{km}- a_{kl} a_{jm}) . &
{}
\end{array}
\eeq
Here the parameter $K$ of deformation involves both $\a$ and Planck's
constant, namely
\beq
          K = \frac{4\a - i h}{4\a + ih},
      \ \ \ \ \ \ \ \ \a^2 = - \frac{1}{3 \L}, \ \ \ \ \ \  \L < 0.
\eeq
Note that in (\ref{clas}) only one copy of the two $SL_{\pm}(2,R)$
is indicated. In conjunction with this, besides the deformed algebra
$A(n)$ derived with, say, $SL_+(2,R)$ taken in (\ref{clas}) and
given by ({\ref{An}}), another identical copy of $A(n)$
(with the only replacement $K\to K^{-1}$) can also be obtained
starting from $SL_-(2,R)$ taken in place of $SL(2,R)$ in (\ref{clas}).
This another copy is independent from the original one:
their generators mutually commute.


\vspace{3mm}
\ni {\large\bf 4. Isomorphism of the 
                  algebras $A(n)$ and $U_q(so_n)$}
\vspace{2mm}

To establish isomorphism between the algebra $A(n)$
from (\ref{An}) and the nonstandard $q$-deformed
algebra $U_q(so_n)$ one has to make the following two steps.

\vspace{3.2mm}
\ \ \ \ --- \ \ { Redefine}:

    \vspace{-9.2mm}
\[  \hspace{34mm}
       \{ K^{1/2} ({K-1})^{-1} \} a_{ik} \longrightarrow  A_{ik},
\]

\ \ \ \ --- \ \ { Identify}:
\vspace{-8.0mm}
\[  \hspace{34mm}
    A_{ik} \longrightarrow  I_{ik},
    \hspace{10mm}
    K \longrightarrow  q .
\]
Then, the Nelson-Regge algebra $A(n)$ is seen to translate
exactly into the nonstandard $q$-deformed algebra $\Uqso$
described above, see (\ref{f2}). We conclude that these two
deformed algebras are isomorphic to each other
(of course, for $K\ne 1$). Recall that $n$ is linked
to the genus $g$ as $\ n=2g+2$, while
$K=(4\a - ih)/(4\a + ih)$ with $\a^2=-\frac{1}{\L}$.

Let us remark that it is the bilinear presentation (2) of the
$q$-algebra $U_q(so_n)$ which makes possible establishing
of this isomorphism. It should be stressed also that
the algebra $A(n)$ plays the role of "intermediate" one:
starting with it and reducing it appropriately,
the algebra of quantum observables (gauge invariant
global characteristics) is to be finally constructed.
The role of Casimir operators in this process, as seen in
\cite{NR}, is of great importance. In this respect let us
mention that the quadratic and higher Casimir elements
of the $q$-algebra $U_q(so_n)$, for $q$ being not a root of $1$,
are known in explicit form \cite{NUW2,GI2} along with eigenvalues
of their corresponding (representation) operators \cite{GI2}.

As shown in detail in \cite{NRZ}, the deformed algebra for the case
of genus $g=1$ surfaces reduces to the desired algebra of
three independent quantum observables which coincides with $A(3)$,
the latter being isomorphic to the Fairlie -- Odesskii algebra
$U_q(so_3)$. The case of $g=2$ is significantly more involved:
here one has to derive, starting with the 15-generator algebra
$A(6)$, the necessary algebra of 6 (independent) quantum observables.
J.Nelson and T.Regge have succeeded \cite{NR3} in constructing such
an algebra. Their construction however is highly nonunique and, what
is more essential, isn't seen to be efficiently extendable to general
situation of $g\geq 3$.


\vspace{3mm}
\ni {\large\bf 5. Outlook }
\vspace{2mm}

Our goal in this note was to attract attention to the isomorphism
of the deformed algebras $A(n)$ from \cite{NR} and the nonstandard
$q$-deformed algebras $\Uqso$ introduced in \cite{GK}). The hope is
that, taking into account a significant amount of the already
existing results concerning diverse aspects of $\Uqso$ (the obtained
various classes of irreducible reprsentations, knowledge of Casimir
operators and their eigenvalues depending on representations, etc.)
we may expect for a further progress concerning construction of the
desired algebra of $6g-6$ independent quantum observables for
space surface of genus $g>2$.

\vspace{3mm}
\ni {\large\bf Acknowledgements}
\vspace{2mm}

This work was supported in part by the CRDF award No. UP1-309.

\end{document}